\documentclass[
reprint,
amsmath,
amssymb,
aps,
prb,
citeautoscript,
floatfix
]{revtex4-1}
\usepackage{natbib}
\newcommand{\angstrom}{\textup{\AA}}
\usepackage{graphicx}
\usepackage{dcolumn}
\usepackage{bm}
\usepackage{lipsum}
\usepackage{multirow}
\usepackage{subcaption}
\begin{document}

\title{Strain Fields and the Electronic Structure of Antiferromagnetic CrN}

\author{Tomas Rojas}

\author{Sergio E. Ulloa}
\email{ulloa@ohio.edu}
\affiliation{%
Department of Physics and Astronomy, and Nanoscale and Quantum Phenomena Institute, Ohio University, Athens, Ohio 45701-2979, USA\\
}%

\date{\today}

\begin{abstract}
We present a theoretical analysis of the role that strain plays on the electronic structure of chromium nitride crystals.  We use LSDA+U calculations
to study the elastic constants, deformation potentials and strain dependence of electron and hole masses near the fundamental gap.  We 
consider the lowest energy antiferromagnetic models believed to describe CrN at low temperatures, and apply strain along different directions.  
We find relatively large deformation potentials for all models, and find increasing gaps for tensile strain along most directions.  Most interestingly, we find
that compressive strains should be able to close the relatively small indirect gap ($\simeq 100$ meV) at moderate amplitudes $\simeq 1.3\%$.
We also find large and anisotropic changes in the effective masses with strain, with principal axes closely related to the magnetic ordering of 
neighboring layers in the antiferromagnet.  
It would be interesting to consider the role that these effects may have on typical film growth on different substrates, and the possibility of monitoring
optical and transport properties of thin films as strain is applied. 
\end{abstract}

\maketitle

\section{\label{sec:level1}INTRODUCTION}

Transition metal nitrides (TMNs) form a large family of materials that play a crucial role in many technological applications. Properties that make them useful include hardness and corrosion resistance, along with interesting electrical properties\cite{Navinsek2001}.

Among these materials, chromium nitride (CrN) is used in coatings for bipolar plates in fuel cells \cite{Jagielski2000, Qiu2013,Lavigne2012}, and as non-carbon support for the oxygen reduction reaction on platinum \cite{Yang2013}.
Most interestingly, this material presents unusual electronic and magnetic properties in crystalline form. These include a phase transition that involves both the magnetic ordering and lattice structure, changing from paramagnetic cubic rock salt structure at high temperatures to antiferromagnetic orthorhombic $P_{nma}$ structure at low temperatures. The N\`{e}el temperature has been reported by various groups to be $T_{N}\simeq 280$K \cite{Gall2002}.

There is, however, some discordance in experiments as to the nature of the electronic characteristics accompanying the structural and magnetic transformation. 
The resistivity $\rho$ at room temperature, has been reported to range from $10^{-4}$ to $10^{2} \, \Omega$/m \cite{Gall2002,Mientus1999,Shih1986}, indicating different degrees of sample polycrystallinity and purity. More controversial is perhaps the resistivity behavior below $T_N$. Some studies report metallic behavior, as $\rho$  increases with temperature \cite{Constantin2004,Inumaru2007,Browne1970}, while others report semiconducting/insulating behavior with the resistivity increasing for decreasing temperature \cite{Gall2002,SubramanyaHerle1997}.
This disagreement in the behavior of $\rho(T<T_N)$ has been attributed to the presence of both dopants and likely nitrogen vacancies that make it difficult to measure the intrinsic transport properties of the CrN thin films \cite{Quintela2009ThermoelectricCrN}.

In an attempt to elucidate the presence of a band gap, Gall {\em et al}.\cite{Gall2002} measured optical and transport properties of single crystal CrN films, and estimated an optical gap of $\approx 0.7 eV$. Also ultraviolet photoelectron spectroscopy (UPS) data suggest that the density of states vanishes at the Fermi energy.
On the other hand a previous study by Herle {\em et al}.\cite{SubramanyaHerle1997} reported a smaller band gap of 90 $meV$, obtained from resistivity data of powder samples.
Recent experimental work has also verified that films grown by molecular beam epitaxy (MBE) exhibit a transition from the paramagnetic to the orthorhombic AFM phase at $T_N \simeq 278 K$, although signs of a cubic phase were also found\cite{Alam2017}.

On the theoretical side, Filipetti {\em et al}. \cite{Filippetti1999} have studied possible magnetic and structural configurations of CrN, concluding that a double antiferromagnetic distorted structure is the lowest energy and most probable configuration, in comparison not only with paramagnetic and ferromagnetic phases but with other antiferromagnetic arrangements. According to their calculations (in the absence of $U$), the double antiferromagnetic phase along the [110] plane ($AFM^2_{[110]}$, see Fig.1) exhibits an enhancement in the number of states near the Fermi energy, but it does not cause the opening of a gap. The formation of this two-layer arrangement is seen as the result of the small energy gain provided by a lattice distortion of about 2$\%$. 
In a subsequent study, the structural transition was understood as the consequence of stress disparity among the Cr atoms that face a plane with same magnetic orientation and those that face the opposite\cite{Filippetti2000}. This `magnetic stress' is seen to drive changes in the spin ordering, and provides a link to the underlying lattice structure \cite{Filippetti2000}.

\begin{figure}[h]
   \centering
    \begin{subfigure}[b]{0.2\textwidth}
        \includegraphics[width=\textwidth]{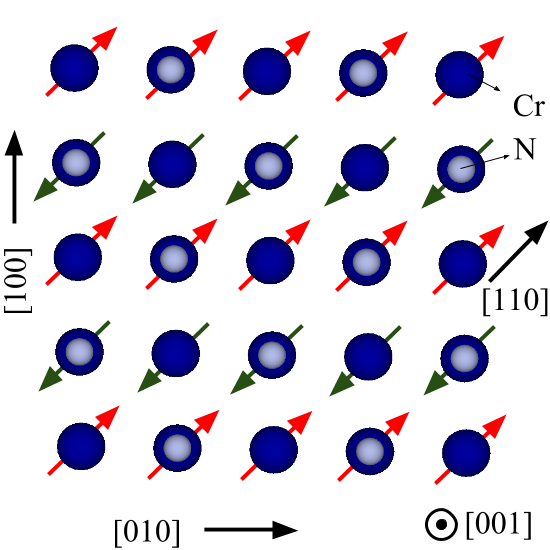}
        \caption{$AFM^1_{[010]}$}
        \label{fig:afm1}
    \end{subfigure}
    ~ 
    \begin{subfigure}[b]{0.2\textwidth}
        \includegraphics[width=\textwidth]{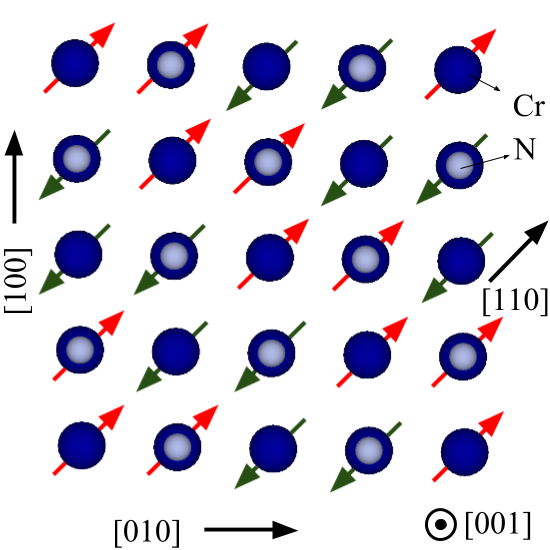}
        \caption{$AFM^2_{[110]}$}
        \label{fig:afm2}
    \end{subfigure}
     \begin{subfigure}[b]{0.2\textwidth}
        \includegraphics[width=\textwidth]{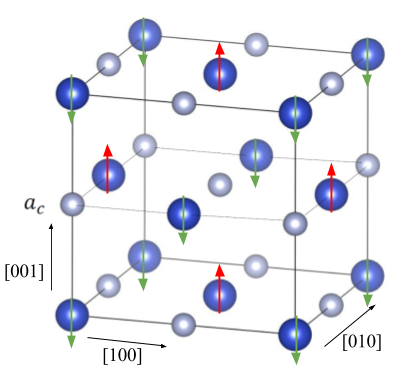}
        \caption{$AFM^1_{[010]}$}
        \label{fig:afm1_bs}
    \end{subfigure}
    ~ 
    \begin{subfigure}[b]{0.23\textwidth}
        \includegraphics[width=\textwidth]{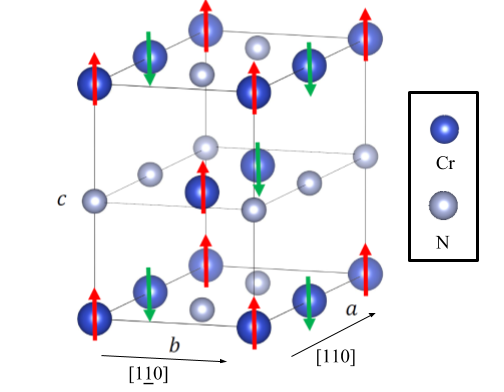}
        \caption{$AFM^2_{[110]}$}
        \label{fig:afm2_bs}
    \end{subfigure}
    \caption{Lattice representation of different AFM models. a) and b) Top view of the $AFM^1_{[010]}$ and $AFM^2_{[110]}$ respectively. c) and d) 3D view of unit cells for cubic $AFM^1_{[010]}$ and orthorhombic $AFM^2_{[110]}$, respectively. Green/red arrows indicate magnetic moment direction on Cr sites and black arrows indicate different lattice directions.}\label{fig:units}
\end{figure}

All these studies suggest that structural and magnetic order are closely linked in CrN and that strains may result in significant changes in magnetic structure.
Indeed, Botana {\em et al}. \cite{Botana2012} carried out a complete computational study of four different configurations, with AFM ordering changing every one or two layers, as well as with and without a $2\%$ distortion. Using a wide variety of functionals, they also conclude that $AFM^2_{[110]}$ is the most stable configuration.
Interestingly, Zhou {\em et al}.\cite{Zhou2014} use an LDA+U approach to show that the energetic difference between the orthorhombic $AFM^2_{[110]}$ and the cubic $AFM^1_{[010]}$ decreases as the value of $U-J$ increases--both models are shown in Fig.\ \ref{fig:units}.
As the reported energy difference between both structures is small, it is conceivable that both phases may appear in a given sample, as perhaps seen recently\cite{Alam2017}.

Recent experimental studies comparing the hardness of the material, after growing it on several substrates, conclude that CrN hardness is independent of the growth orientation and of grain boundary effects\cite{Ikeyama2016a}. Moreover, theoretical studies of strained free-standing CrN films explored the stability of  mono and bilayers\cite{Zhang2014a}. Their results suggest that a biaxial strain of $-6\%$ would cause a phase transition to a ferromagnetic phase at low temperatures for a (100) plane of the material. 
Although CrN films have been grown on a variety of substrates, the most commonly used is MgO; the CrN lattice mismatch with this substrate could be 
$\simeq 1.7\%$, as estimated from the reported lattice constants\cite{Ikeyama2016a}. As film growth is often conducted at elevated temperatures, cooling down 
may induce additional strains due to different thermal expansion coefficients of film and substrate. This suggests that a strain range of $\pm2\%$ could be 
realistically achieved in the growth conditions of CrN films. Higher strains may require externally applied stress, as achieved by film deposition on flexible substrates. \cite{BolotinStrain2013}

As lattice distortion and magnetic behavior are clearly interconnected in this material, we have undertaken a study of electronic properties of CrN under strain.
These could be explored by epitaxial growth of thin films on different substrates and explored by both bulk and surface probes. 
Utilizing an LSDA+U approach, we calculate elastic constants and deformation potentials for different expected antiferromagnetic phases of the material, and explore the changes in band structure the strains produce. We obtain the effective masses and anisotropy for both electron and hole bands near the optical gap, and find that a $1\%$ strain can boost the carrier mass and mobility by a factor of 2 in some directions, while remaining unchanged in others. Moreover, we find that strains in a range of $2\%$ may have other important effects, including the closing of the fundamental (indirect) energy gap for a compressive strain 
$\simeq -1.3\%$ in all the AFM models tested. Apart from providing specific testable predictions for deformation potentials in this material, 
these results may contribute to explaining how some experimental setups have shown metallic behavior for $T<T_N$ and suggest additional experiments to probe the electrical properties of the material.  

\section{\label{sec:level1c}Computational Approach}
We performed density functional calculations in the local spin density approximation (LSDA), implemented in the Quantum Espresso package\cite{Giannozzi2009}. 
We used an LSDA+U approach on the $3d$ partially localized Cr orbitals of the system\cite{Anisimov1997}, in a rotationally invariant formulation\cite{Liechtenstein1995}. The kinetic energy cutoff for the wavefunctions was set to $30$ Ry and the Brillouin zone was sampled with $8\times 8 \times 8$ k-point meshes.
An important aspect in using LSDA+U is the validation of estimates of the $ad$ $hoc$ constants $U$ and $J$ in the implementation; we take advantage of previous work on this material. Herwadkar {\em et al}.\cite{Herwadkar2009} considered both an excited-atom model in which the screening charge is restricted to a sphere where $d$ electrons can be added or removed, and the Cococcione and Gironcoli algorithm\cite{Cococcioni2005} to estimate reasonable values for the $U$ and $J$ parameters. They conclude that a reasonable value of $U$ is in the range 3 to 5 eV, and that $J=0.94$ eV. Similar values of $U-J$ are used by Zhou {\em et al}.\cite{Zhou2014}
With this input, we performed electronic structure calculations on the low energy models in Fig.\ \ref{fig:units} and consider distortion effects. We use $J=0.94$ eV and $U=3$ eV for the $AFM^2_{[110]}$ model, but $U=5$ eV for $AFM^1_{[010]}$, to make both models have a comparable band gap. We notice that other $U$ values shift the final energy gap but cause no major difference in the electronic spectrum for each model; similarly the deformation
potentials and effective masses presented here are unaffected over the range of anticipated $U-J$ values.
As in previous computational studies\cite{Botana2012,Zhou2014,Filippetti1999}, the orthorhombic $AFM^2_{[110]}$ antiferromagnetic model was found to have the lowest energy, followed by the cubic $AFM^2_{[110]}$ at $8.0$ meV per chemical unit. The cubic $AFM^1_{[010]}$ was found to be at $15$ meV from the minimum when the same set of $U$ and $J$ parameters are used. Finally, the orthorhombic $AFM^1_{[010]}$ was found at $30$ meV.

\begin{figure}[h]
        \includegraphics[scale=0.54]{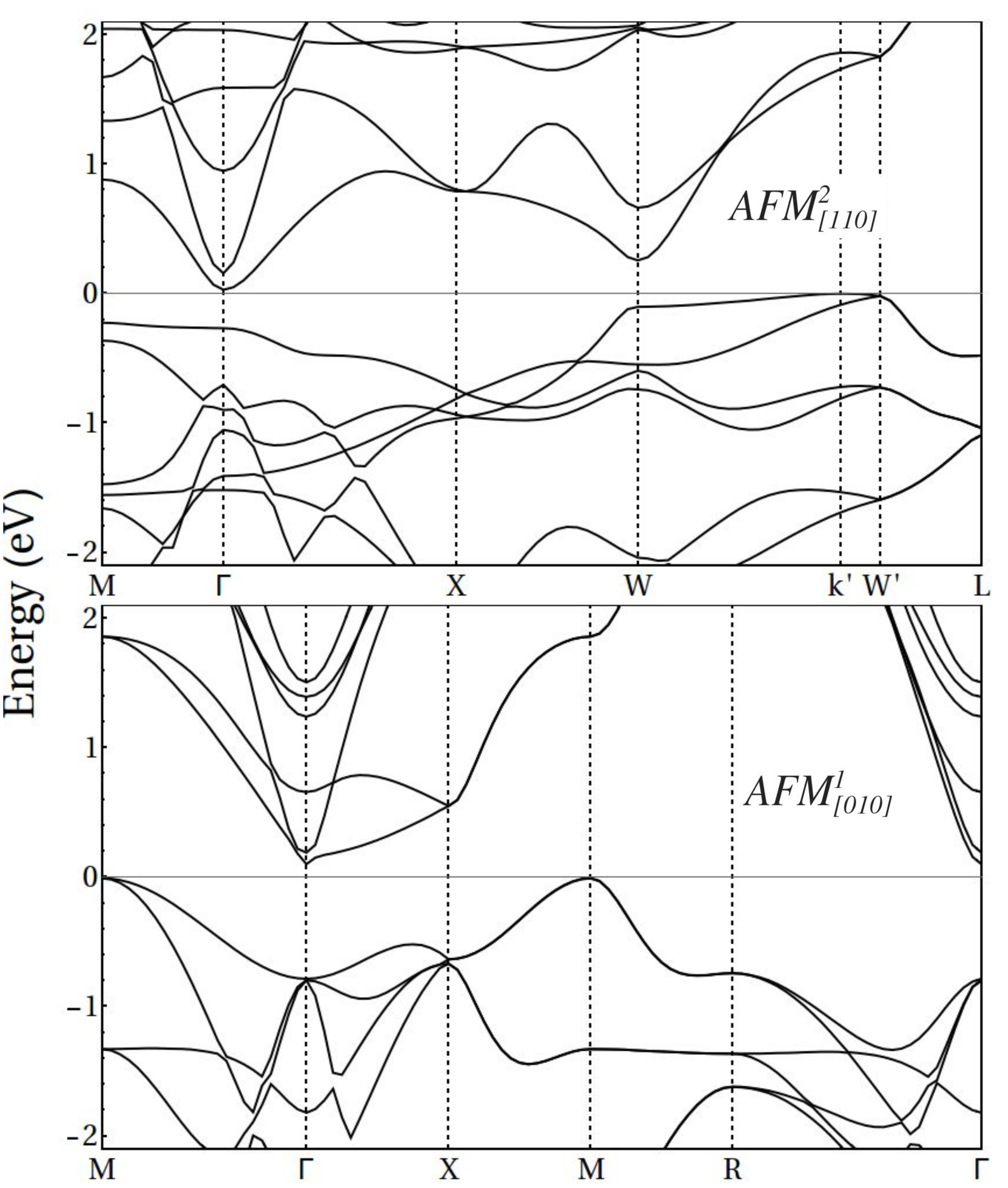}
    \caption{Low energy AFM band structures for models of CrN. Top: Orthorhombic $AFM^2_{[110]}$, and Bottom: Cubic $AFM^1_{[010]}$.}
    \label{fig:bands}
\end{figure}

Figure \ref{fig:bands} shows band structures for both main models. The symmetry points shown for the cubic $AFM^1_{[010]}$ are $\Gamma=(0,0,0)$, $M=(\frac{1}{2},\frac{1}{2},0)2\pi/a_c$, $X=(\frac{1}{2},0,0)2\pi/a_c$ and $R=(\frac{1}{2},\frac{1}{2},\frac{1}{2})2\pi/a_c$, where $a_c=4.14 \angstrom$ is the cubic lattice constant. For the orthorhombic $AFM^2_{[110]}$, the symmetry points correspond to $M=(\pi/a,0,0)$, $X=(0,\pi/b,0)$, $W=(\pi/a,3\pi/2b,0)$, $W'=(\pi/a,3\pi/2b,\pi/c)$ and $L=(0,\pi/b,\pi/a_c)$, with $a=2 a_c \sin(\alpha /2)$ and $b=a_c \cos(\alpha/2)$, where $\alpha = 88^{\circ}$. In the $AFM^2_{[110]}$ structure the valence band maximum at $k'$ is away from the symmetry point (see Fig.\ \ref{fig:bands}), surpassing the $W$ point by $\simeq 10$ meV, as mentioned in the literature.\cite{Botana2012}
An interesting difference among the models are the orbitals responsible for the energy gap; the $AFM^{1}_{[010]}$ shows a hybridization gap between $d_{z^2}$ in the conduction band at the $\Gamma$ point, and $p_z$ in the valence at $M$. In contrast, in the $AFM^{2}_{[110]}$ structure the conduction band has $d_{z^2}$ character at $\Gamma$, while $d_{x^2-y^2}$ dominates in the valence band at $W$.

As we will see below, the nature of the orbitals involved in the indirect gap is reflected in the behavior of the band gap variation with various strains, as well as the symmetries and strain changes produced on the different effective masses.

\section{\label{sec:level1r}Results}
We calculate the effects of strain in the band structure of different $AFM$ model structures, as described before, and for different uniaxial directions of strain:  $[110]$, $[1\bar{1}0]$, $[010]$, and $[100]$, as well as for biaxial strains. For each strain amplitude, the unit cell is stretched along the given direction and then allowed to relax in the remaining directions.
Strains are naturally found to shift bands and change their curvature. As such, we have carried out an analysis of these effects considering the two lowest energy models for the AFM low temperature phase, the orthorhombic $AFM^2_{[110]}$ and the cubic $AFM^1_{[010]}$. Notice, however, that recent RHEED analyses of epitaxial films suggest the possibility of a cubic $AFM^2_{[110]}$ phase\cite{Alam2017}, which is also found to have relatively low configuration energy, as discussed above. This has motivated us to also analyze the behavior of such cubic $AFM^2_{[110]}$ phase, as well as the corresponding orthorhombic $AFM^ 1_{[010]}$ phase, for completeness. If these phases are in fact present in films, it would be interesting to explore their behavior under possible strain fields, which may contribute to their identification.

\subsection{\label{sec:level2ec}Elastic Constants}
We have calculated the elastic constants $C_{11}$, $C_{12}$ and $C_{44}$. The first two are obtained from the bulk $B$ and shear modulus $C_s$, using the expressions $B = (C_{11} + 2C_{12})/3$ and $C_s = (C_{11} - C_{12})/2$\cite{Wright1997}. 
The bulk modulus is obtained by applying hydrostatic pressure and fitting the energy and lattice constant relation to the Murnaghan equation of state \cite{Murnaghan1944}.
The shear modulus $C_s$ and $C_{44}$ are obtained by applying volume-conserving tetragonal and trigonal strains respectively, as done previously in other nitrides\cite{Wright1997}.
For the cubic $AFM^1_{[010]}$ model, we obtain $C_{11}=677$ GPa, $C_{12}=87.7$ GPa, $C_{44}=155$ GPa and  $B=289$ GPa, in general agreement with previous calculations\cite{Antonov2010,Liang2010,JIAO2013} and experiments\cite{Almer2003,Martinschitz2009}. A full discussion of the role of DFT functionals on different CrN elastic constants and structures is presented by Zhou {\em et al}.\cite{Zhou2014} We notice that these large elastic constants show some proportionality with the antiferromagnetic phase of a chromium crystal\cite{Muir1987}, where $C_{11}=428$ GPa, $C_{12}=51$ GPa and $C_{44}=96$ GPa. There is a factor of 4 between $C_{44}$ and $C_{11}$ and a factor of about 8 between $C_{12}$ and $C_{11}$. In contrast, other non-magnetic nitrides\cite{Wright1997} such as GaN ($C_{11}=293$ GPa, $C_{12}=159$ GPa and $C_{44}=155$ GPa) and AlN ($C_{11}=304$ GPa, $160$ GPa and $C_{44}=193$ GPa) are softer and have values of $C_{44}$ and $C_{12}$ close to each other, and only a factor of 2 smaller than $C_{11}$.

\subsection{\label{sec:level2eg}Energy Gaps and Deformation Potentials}
\begin{figure}[t]
   \centering \includegraphics[scale=0.23]{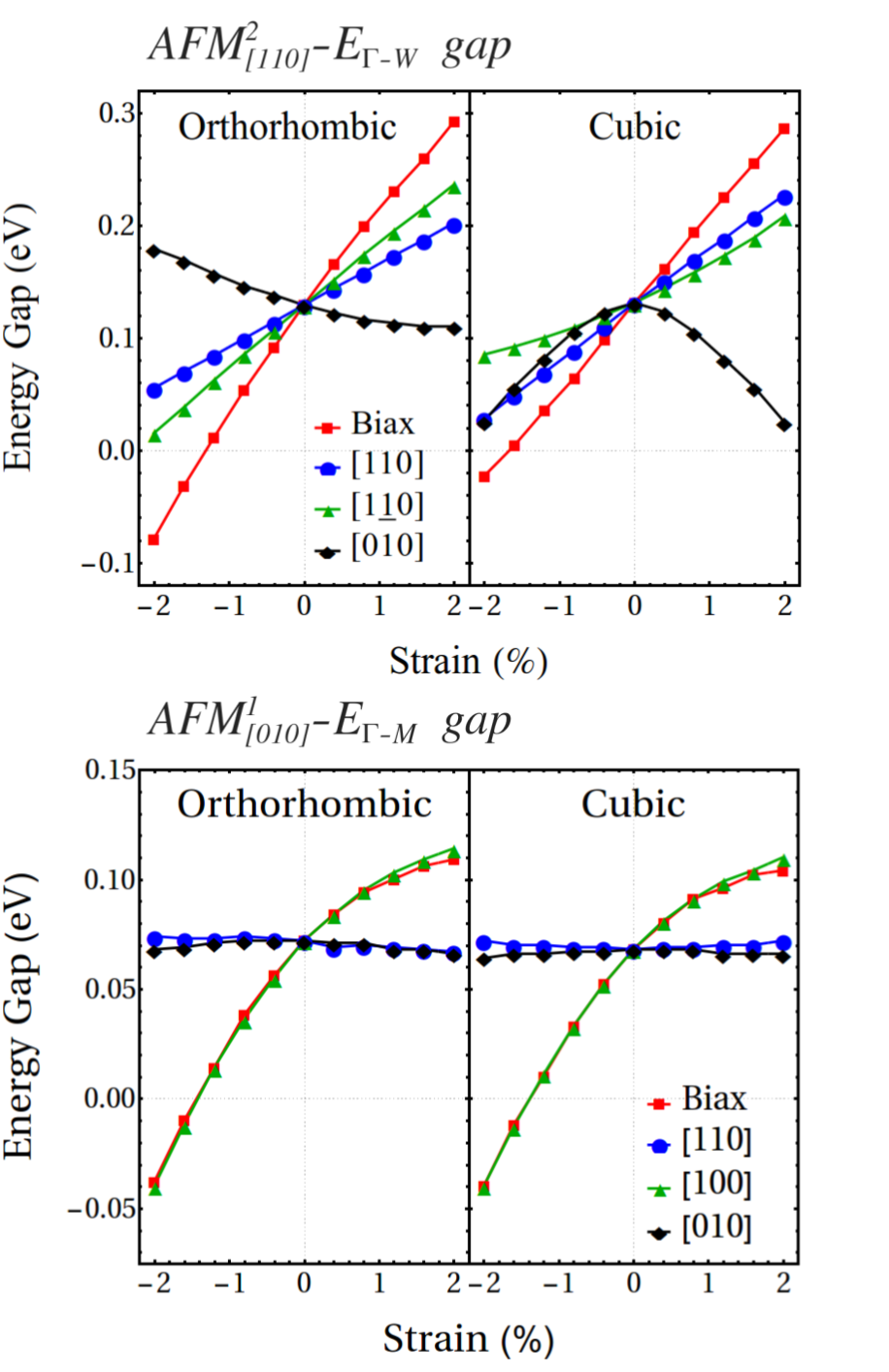}
	\caption{Indirect energy gap dependence with strain for different strain axes and AFM models of CrN. Top panel: $AFM^2_{[110]}$. Bottom panel: $AFM^1_{[010]}$.}
	\label{fig:gaps}
\end{figure}

Figure \ref{fig:gaps} shows the change in the indirect gaps with strain ($E_{\Gamma-M}$ or $E_{\Gamma-W}$) for all models. In orthorhombic $AFM^2_{[110]}$, compressive strains along the AFM axis ($[110]$) reduce only slightly the $E_{\Gamma-W}$ gap with a deformation potential $D_{[110]}=\frac{dE_{gap}}{d\epsilon}=3.65$ eV, and such near linear dependence continues for tensile strain. This is similar, although slightly larger for strains along $[1\bar{1}0]$, 
with $D_{[1\bar{1}0]}=5.52$ eV, and nearly additive for 
biaxial ([110]+[$1\bar{1}0$]) strain, with $D_{\rm biax}=9.17$ eV.\@
Along the cubic axes $[010]$ and $[100]$, there is an opposite linear tendency with $D_{[010]} \simeq D_{[100]} \simeq -1.80$ eV, decreasing slightly for
tensile strains. We find that similar tendencies remain true in the cubic $AFM^2_{[110]}$, where for the $[110]$, $[1\bar{1}0]$ and biaxial strain, we find $D_{[110]}=4.98$ eV, $D_{[1 \bar{1}0]}=3.05$ eV and $D_{\rm biax}=7.85$ eV.\@  However, for the $[010]$ strain we find that both compressive and tensile strains decrease the band gap at a rate of $\simeq \pm 5.5$ eV, with nearly quadratic dependence.

The indirect gap $E_{\Gamma-M}$ for cubic $AFM^1_{[010]}$ behaves quite differently to the gap in $AFM^2_{[110]}$. 
For this case, the gap remains nearly unchanged for the $[010]$, $[110]$ and $[1\bar{1}0]$ directions (last two equivalent in this cubic model). 
On the other hand, for strains along $[100]$ and biaxial ([010]+[100]), the gaps change  at a similar rate of $D_{[010]}=3.54$ eV and $D_{\rm biax}=3.67$ eV,
decreasing slightly for tensile strains. For the distorted orthorhombic $AFM^1_{[010]}$, the deformation potentials are nearly the same as for the
cubic model, as seen in the bottom panel of Fig.\ \ref{fig:gaps}.

It is interesting that the values for deformation potentials are substantially larger than in materials sharing similar crystalline structure, such as ScN, where 
the $D$s are in the range of $1.36-2.06$ eV \cite{Qteish2006}.  The values here are comparable to those of Si for strain along the $[110]$ direction\cite{Dhar2007}. We also notice that in all of these materials the energy gap increases with tensile strain and decreases with compression, except for the orthorhombic model $AFM^2_{[110]}$ for strains along [010].
In materials like GaN and AlN, the deformation potentials for hydrostatic strain are in fact negative, $-8.31$ eV and $-9.88$ eV, respectively\cite{Kim1997},
with large (and negative) uniaxial deformation potentials. \cite{QinStrains} 

\begin{figure}[t]
    \centering \includegraphics[scale=0.23]{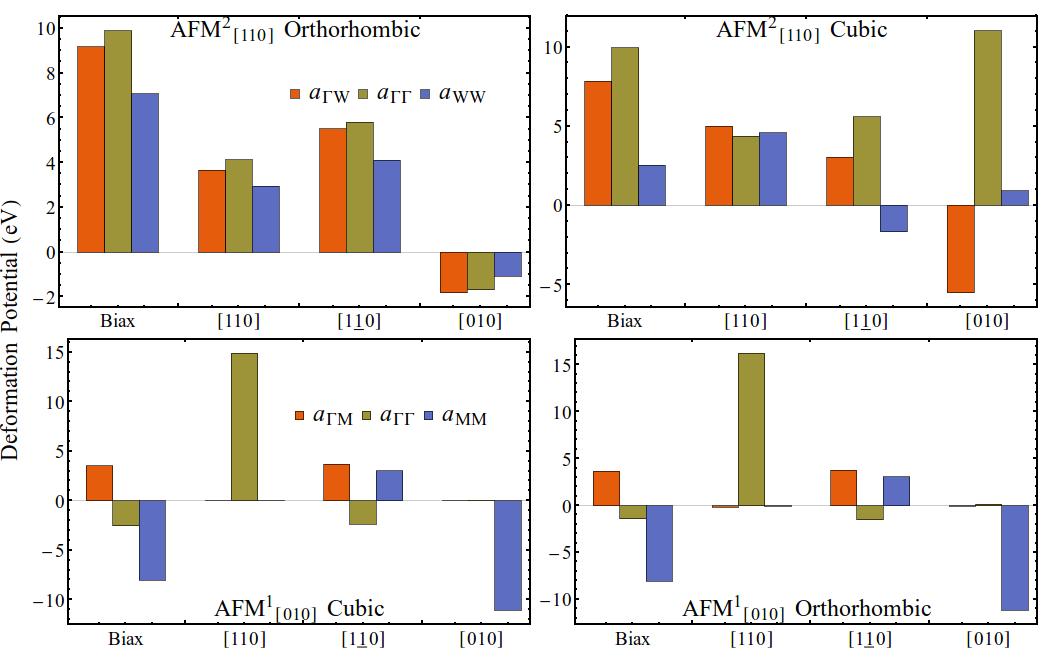}
    \caption{Deformation potentials for all AFM models considered, for both direct and indirect gaps, as indicated.}
    \label{fig:defpot}
\end{figure}

A summary of deformation potentials for both direct and indirect band gaps in all the four AFM models is shown in Fig.\ \ref{fig:defpot}. Notice the deformation potentials decrease slightly for the distorted structures, but maintain similar tendencies, with the exception of $[010]$ in the $AFM^2_{[110]}$.

An interesting consequence of the relatively large deformation potentials seen in Fig.\ 3 and 4, is that compressive strains reduce strongly the fundamental (indirect) gap in the band structure. In fact, for the orthorhombic $AFM^2_{[110]}$ phase, our calculations show that the gap closes for a biaxial strain $\approx -1.3\%$, while a uniaxial strain would require $\approx -2.2\%$. Similarly, the cubic $AFM^1_{[010]}$ structure would be gapless for both biaxial and $[100]$ uniaxial strains $\approx -1.4\%$. As this level of strains ($\lesssim 2\%$) is likely achievable by appropriate choice of epitaxial substrate and/or by differential expansion coefficients, it may be interesting to monitor the strain field in films while monitoring their transport and/or optical properties. The reduction of the gap in conjunction with the AFM ordering may give rise to interesting piezomagnetic behavior in this material.  We should note that a larger value of the initial gap (as obtained with a larger $U$, for example), would also require larger compressive strains to close.  We anticipate, however, that in all cases the required strain would be $\lesssim 2\%$.

\subsection{\label{sec:level2em}Effective masses}
Strains are also expected to change the curvature of bands. To estimate the effective masses and their strain dependence, band surfaces of valence and conduction bands were calculated using a mesh $300$ k-points around the high symmetry points, and then fitted to the effective mass tensor,
\begin{equation}
m^{-1}= \frac{1}{\hbar^2}
\begin{pmatrix}
\frac{\partial^2 E}{\partial k^2_{x}}&\frac{\partial^2 E}{\partial k_{x}\partial k_{y}}&\frac{\partial^2 E}{\partial k_{x}\partial k_{z}}\\
\frac{\partial^2 E}{\partial k_{y}\partial k_{x}}&\frac{\partial^2 E}{\partial k^2_{y}}&\frac{\partial^2 E}{\partial k_{y}\partial k_{z}}\\
\frac{\partial^2 E}{\partial k_{z}\partial k_{x}}&\frac{\partial^2 E}{\partial k_{z}\partial k_{y}}&\frac{\partial^2 E}{\partial k^2_{z}}\\
\end{pmatrix}.
\end{equation}

\begin{figure}
    \centering \includegraphics[scale=0.46]{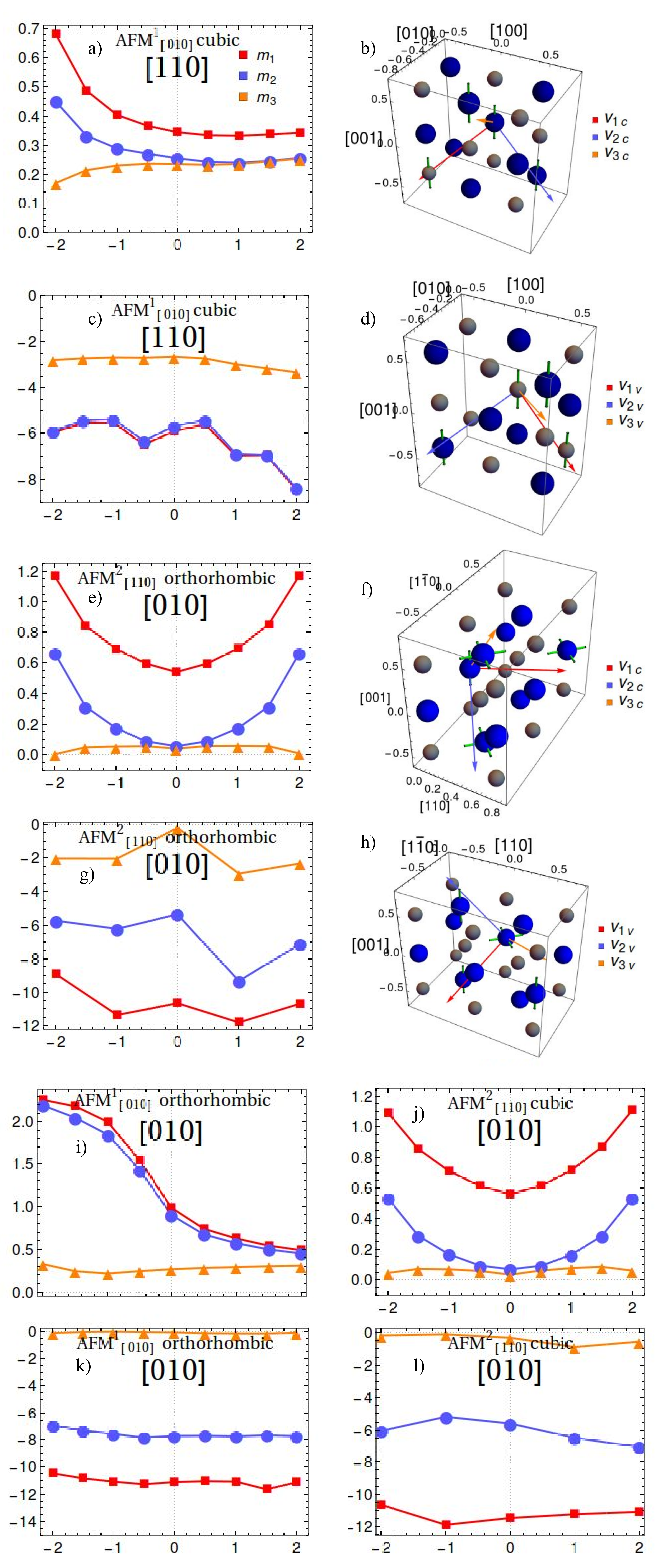}
    \caption{a) Electron and c) Hole effective masses for the  cubic $AFM^1_{[010]}$ model with $[110]$ strains. b) and d) Show corresponding principal axes for masses, as indicated by color coding in a) and c). Green rods through atoms indicate $d_{z^2}$ (for Cr) or $p_z$ (for N) orbitals for reference. e) and g) Effective masses for the orthorhombic $AFM^2_{[110]}$ model for $[010]$ strains. f) and h) Principal axes for masses in e) and g). i) and k) Effective masses for the  orthorhombic $AFM^1_{[010]}$ model with $[010]$ strains. j) and l) Effective mass for the  cubic $AFM^2_{[110]}$ model with $[010]$ strains.}
    \label{fig:mass}
\end{figure}

The fitted tensor accurately describes dispersion up to at least $0.4$ eV away from each band edge. The resulting asymmetric tensor is diagonalized to obtain the masses along the principal axes. 
All AFM models exhibit their principal axes not fully aligned with the cubic directions, and are rather better aligned with directions to the nearest neighbor 
atoms in the unit cell, as shown for a few examples in Fig.\ \ref{fig:mass}. 
Specifically, in the cubic $AFM^1_{[010]}$ model, the conduction band (Fig.\ \ref{fig:mass}a and b) shows two degenerate masses of $0.24 m_e$ with principal vectors directed towards the nearest neighbor Cr atoms in a different layer and therefore with different spin in the AFM configuration ($m_e$ is the bare electron mass). Those masses are smaller than the remaining one (in red) with $0.34 m_e$, with principal axis pointing towards the nearest N atom. Interestingly, the masses are rather insensitive to tensile strains, while changing drastically under $[110]$ compression. Likewise, the valence band (Fig.\ \ref{fig:mass}c and d) has two equal masses of $-5.9 m_e$ with principal axes pointing from an N atom to the Cr atoms in the next plane parallel to $[010]$. The smaller
mass $\simeq -2.8m_e$ shows weaker strain dependence, although all hole masses increase with strain. 

In contrast, in the orthorhombic $AFM^2_{[110]}$, the conduction band (Fig.\ \ref{fig:mass}e and f) has two degenerate small masses of $0.05 m_e$, with principal axes in the directions of the Cr atoms in the next plane parallel to $[110]$, and therefore with the same spin. The remaining mass of $0.55 m_e$ has an axis pointing to the Cr atom in the second nearest plane parallel to $[110]$ which correspond to a different spin. 
The lower symmetry of this structure is also reflected in the rather different and generally larger masses in the valence band, Fig.\ \ref{fig:mass}g and h.

One can intuitively interpret the principal mass directions as connected to the interaction between the orbitals present near the corresponding energy for each case. For instance, in the $AFM^1_{[010]}$, the axes are clearly influenced by the position of the density lobes of the $p_z$ and $d_{z^2}$ orbitals (shown as green rods in Fig.\ \ref{fig:mass}). 

Although, Fig.\ \ref{fig:mass} shows only a subset of all possible results, we can describe the general behavior of masses with strain.
As mentioned, the impact of the up to $\pm 2 \%$ distortions on the various masses is different. 
The mass changes with strain have different dependence in conduction and valence bands. The valence bands show typically more pronounced changes 
for tensile strain in the $AFM^2_{[110]}$ structure.  In this structure, strain along $[110]$ produces small changes in all the masses, changing less than $8\%$ with a $2\%$ strain. On the other hand, strains along $[1\bar{1}0]$ changes masses up to $60 \%$ in the valence band and up to $35\%$ in the conduction band (not shown). 
It is also interesting to notice the general symmetry breaking between mass pairs in the conduction band, as strains set in. 
Lastly, the biaxial strain produces the largest changes, with valence band masses decreasing in most cases by up to $60 \%$ at $2\%$ strain. In the conduction band,  masses change by about $40 \%$ for similar strains. 

The $AFM^1_{[010]}$ structure shows overall smaller changes in comparison, particularly for $[010]$ strains.
One finds a change of about $25\%$ in the valence band and less than $5\%$ in the conduction band masses. $[100]$ strains produce a symmetry breaking in the conduction band for $m_2$ and $m_3$, with changes of about $14\%$. Similar results are present for the $[110]$ strain (Fig.\ \ref{fig:mass}a) with changes of up to $47 \%$ in the valence band and degeneracy of $m_2$ and $m_3$ broken for compressive strain, with changes of up to $80\%$. Finally, biaxial strains show smaller changes in comparison, with $22\%$ changes in the valence band and $15\%$ in the conduction band masses. 

We have also studied the distorted counterparts of the models. Particularly, the conduction band of the orthorhombic $AFM^1_{[010]}$ model for the $[010]$ strain shows increases of $~100\%$ in $m_1$ and $m_2$, with no change for $m_3$, as seen in Fig.\ \ref{fig:mass}i.  There is, in contrast, nearly 
no change in the valence band masses, as seen in Fig.\ \ref{fig:mass}k. The degeneracy of $m_1$ and $m_2$ is preserved for all strains in this model, 
unlike in the cubic $AFM^1_{[010]}$. The biaxial strain shows almost identical changes in the effective masses to the cubic case, while $[110]$ 
strains produce similar percentual changes but with a more quadratic relation (not shown). Strains along $[100]$ show 
opposite tendency, with $m_1$ and $m_2$ increasing only under tensile strain by $\approx 50\%$, with no change for $m_3$ in this strain range.
In the cubic $AFM^2_{[110]}$ model, $[010]$ strains result in almost identical changes in the conduction band to that of the orthorhombic $AFM^2_{[110]}$ model (Fig.\ \ref{fig:mass}j), with significantly smaller change in the valence band masses ($\lesssim 2\%$), as seen in Fig.\ \ref{fig:mass}l. 
Both strains $[110]$ and $[1\bar{1}0]$ produce smaller changes ($\approx 10\%$) in all effective masses, while biaxial strain splits the degeneracy between $m_2$ and $m_1$, doubling $m_2$ but increasing $m_1$ only about $40\%$ in the conduction band for tensile strains; compressive strains result in
nearly no change (not shown). In the valence band, strain produces a nearly linear change of $m_2$ and $m_3$ of $\lesssim 20\%$, 
and no significant change in $m_1$.

It is clear that all changes of effective mass with strain are rather anisotropic, as we have seen. This would likely result in corresponding transport coefficients that inherit such asymmetries. Monitoring of transport properties with changing strain in doped samples (naturally or purposely) would provide interesting insights into the behavior of this material. 

\section{\label{sec:conclusions} Conclusions}
We have studied the effects that moderate strains of $\pm2\%$ have in band gaps, deformation potentials and effective masses of different low temperature phases of CrN, orthorhombic $AFM^2_{[110]}$ and cubic $AFM^1_{[010]}$, as well as their corresponding low-energy distortions.
Our results show that compressive biaxial and uniaxial strains $\leq 2\%$ are able to close the gap in these AFM  structures. This suggests that realistic strain could cause significant changes in the electronic structure and optical response, and could contribute to explaining some of the puzzling behavior seen under different experimental conditions and film growth characteristics. Moreover, these results suggest experiments to closely monitor transport and/or optical response of CrN films as applied strains are varied. 
The transport properties in doped systems will also reflect changes in the effective masses estimated by our calculations. As masses show large anisotropy, different directions of strain would also produce different reactions on the charge carrier mobility. 
The significant changes in charge carrier mobility that a strain of $\pm 2\%$ causes could be a factor that needs to be considered in the appearance of the resistivity kink seen to occur in several studies\cite{Inumaru2007,Constantin2004,Bhobe2010}. 
It would be interesting to probe to
what extent this could be part of a strain relief mechanism in different sample conditions and morphology. 
Another aspect to consider is that although the elastic constants hardly differ among the $AFM^2_{[110]}$ and $AFM^1_{[010]}$ structures, the deformation potentials as well as the effective mass changes are not at all similar. This behavior could be the result of the magnetic stress described by Fillipetti {\em et al}.\cite{Filippetti2000} and appearing due to the different stress experienced by the various parallel magnetic layers of the material, present in $AFM^2_{[110]}$ but absent in $AFM^1_{[010]}$.

\section*{\label{sec:ack} Acknowledgments}
We thank Walter Lambrecht, Arthur Smith, and Khan Alam for helpful and interesting discussions on this material.
The work was supported by NSF grant DMR-1508325. Most calculations were performed at the Ohio Supercomputing Center under project PHS0219.
\bibliographystyle{apsrev4-1}
\bibliography{MendeleyShort.bib}

\begin{thebibliography}{39}%
\makeatletter
\providecommand \@ifxundefined [1]{%
 \@ifx{#1\undefined}
}%
\providecommand \@ifnum [1]{%
 \ifnum #1\expandafter \@firstoftwo
 \else \expandafter \@secondoftwo
 \fi
}%
\providecommand \@ifx [1]{%
 \ifx #1\expandafter \@firstoftwo
 \else \expandafter \@secondoftwo
 \fi
}%
\providecommand \natexlab [1]{#1}%
\providecommand \enquote  [1]{``#1''}%
\providecommand \bibnamefont  [1]{#1}%
\providecommand \bibfnamefont [1]{#1}%
\providecommand \citenamefont [1]{#1}%
\providecommand \href@noop [0]{\@secondoftwo}%
\providecommand \href [0]{\begingroup \@sanitize@url \@href}%
\providecommand \@href[1]{\@@startlink{#1}\@@href}%
\providecommand \@@href[1]{\endgroup#1\@@endlink}%
\providecommand \@sanitize@url [0]{\catcode `\\12\catcode `\$12\catcode
  `\&12\catcode `\#12\catcode `\^12\catcode `\_12\catcode `\%12\relax}%
\providecommand \@@startlink[1]{}%
\providecommand \@@endlink[0]{}%
\providecommand \url  [0]{\begingroup\@sanitize@url \@url }%
\providecommand \@url [1]{\endgroup\@href {#1}{\urlprefix }}%
\providecommand \urlprefix  [0]{URL }%
\providecommand \Eprint [0]{\href }%
\providecommand \doibase [0]{http://dx.doi.org/}%
\providecommand \selectlanguage [0]{\@gobble}%
\providecommand \bibinfo  [0]{\@secondoftwo}%
\providecommand \bibfield  [0]{\@secondoftwo}%
\providecommand \translation [1]{[#1]}%
\providecommand \BibitemOpen [0]{}%
\providecommand \bibitemStop [0]{}%
\providecommand \bibitemNoStop [0]{.\EOS\space}%
\providecommand \EOS [0]{\spacefactor3000\relax}%
\providecommand \BibitemShut  [1]{\csname bibitem#1\endcsname}%
\let\auto@bib@innerbib\@empty
\bibitem [{\citenamefont {Navinsek}\ and\ \citenamefont
  {Seal}(2001)}]{Navinsek2001}%
  \BibitemOpen
  \bibfield  {author} {\bibinfo {author} {\bibfnamefont {B.}~\bibnamefont
  {Navinsek}}\ and\ \bibinfo {author} {\bibfnamefont {S.}~\bibnamefont
  {Seal}},\ }\href {\doibase 10.1007/s11837-001-0072-1} {\bibfield  {journal}
  {\bibinfo  {journal} {{JOM}}\ }\textbf {\bibinfo {volume} {53}},\ \bibinfo
  {pages} {51} (\bibinfo {year} {2001})}\BibitemShut {NoStop}%
\bibitem [{\citenamefont {Jagielski}\ \emph {et~al.}(2000)\citenamefont
  {Jagielski}, \citenamefont {Khanna}, \citenamefont {Kucinski}, \citenamefont
  {Mishra}, \citenamefont {Racolta}, \citenamefont {Sioshansi}, \citenamefont
  {Tobin}, \citenamefont {Thereska}, \citenamefont {Uglov}, \citenamefont
  {Vilaithong}, \citenamefont {Viviente}, \citenamefont {Yang},\ and\
  \citenamefont {Zalar}}]{Jagielski2000}%
  \BibitemOpen
  \bibfield  {author} {\bibinfo {author} {\bibfnamefont {J.}~\bibnamefont
  {Jagielski}}, \bibinfo {author} {\bibfnamefont {A.}~\bibnamefont {Khanna}},
  \bibinfo {author} {\bibfnamefont {J.}~\bibnamefont {Kucinski}}, \bibinfo
  {author} {\bibfnamefont {D.}~\bibnamefont {Mishra}}, \bibinfo {author}
  {\bibfnamefont {P.}~\bibnamefont {Racolta}}, \bibinfo {author} {\bibfnamefont
  {P.}~\bibnamefont {Sioshansi}}, \bibinfo {author} {\bibfnamefont
  {E.}~\bibnamefont {Tobin}}, \bibinfo {author} {\bibfnamefont
  {J.}~\bibnamefont {Thereska}}, \bibinfo {author} {\bibfnamefont
  {V.}~\bibnamefont {Uglov}}, \bibinfo {author} {\bibfnamefont
  {T.}~\bibnamefont {Vilaithong}}, \bibinfo {author} {\bibfnamefont
  {J.}~\bibnamefont {Viviente}}, \bibinfo {author} {\bibfnamefont {S.-Z.}\
  \bibnamefont {Yang}}, \ and\ \bibinfo {author} {\bibfnamefont
  {A.}~\bibnamefont {Zalar}},\ }\href {\doibase 10.1016/S0169-4332(99)00350-5}
  {\bibfield  {journal} {\bibinfo  {journal} {Appl. Surf. Sci.}\ }\textbf
  {\bibinfo {volume} {156}},\ \bibinfo {pages} {47} (\bibinfo {year}
  {2000})}\BibitemShut {NoStop}%
\bibitem [{\citenamefont {Qiu}\ \emph {et~al.}(2013)\citenamefont {Qiu},
  \citenamefont {Zhang}, \citenamefont {Li}, \citenamefont {Wang},
  \citenamefont {Lee}, \citenamefont {Li},\ and\ \citenamefont
  {Zhao}}]{Qiu2013}%
  \BibitemOpen
  \bibfield  {author} {\bibinfo {author} {\bibfnamefont {Y.}~\bibnamefont
  {Qiu}}, \bibinfo {author} {\bibfnamefont {S.}~\bibnamefont {Zhang}}, \bibinfo
  {author} {\bibfnamefont {B.}~\bibnamefont {Li}}, \bibinfo {author}
  {\bibfnamefont {Y.}~\bibnamefont {Wang}}, \bibinfo {author} {\bibfnamefont
  {J.-W.}\ \bibnamefont {Lee}}, \bibinfo {author} {\bibfnamefont
  {F.}~\bibnamefont {Li}}, \ and\ \bibinfo {author} {\bibfnamefont
  {D.}~\bibnamefont {Zhao}},\ }\href {\doibase 10.1016/j.surfcoat.2012.03.010}
  {\bibfield  {journal} {\bibinfo  {journal} {Surf. Coat. Tech.}\ }\textbf
  {\bibinfo {volume} {231}},\ \bibinfo {pages} {357} (\bibinfo {year}
  {2013})}\BibitemShut {NoStop}%
\bibitem [{\citenamefont {Lavigne}\ \emph {et~al.}(2012)\citenamefont
  {Lavigne}, \citenamefont {Alemany-Dumont}, \citenamefont {Normand},
  \citenamefont {Berthon-Fabry},\ and\ \citenamefont
  {Metkemeijer}}]{Lavigne2012}%
  \BibitemOpen
  \bibfield  {author} {\bibinfo {author} {\bibfnamefont {O.}~\bibnamefont
  {Lavigne}}, \bibinfo {author} {\bibfnamefont {C.}~\bibnamefont
  {Alemany-Dumont}}, \bibinfo {author} {\bibfnamefont {B.}~\bibnamefont
  {Normand}}, \bibinfo {author} {\bibfnamefont {S.}~\bibnamefont
  {Berthon-Fabry}}, \ and\ \bibinfo {author} {\bibfnamefont {R.}~\bibnamefont
  {Metkemeijer}},\ }\href {\doibase 10.1016/j.ijhydene.2012.04.035} {\bibfield
  {journal} {\bibinfo  {journal} {Int. J. Hydrogen Energ.}\ }\textbf {\bibinfo
  {volume} {37}},\ \bibinfo {pages} {10789} (\bibinfo {year}
  {2012})}\BibitemShut {NoStop}%
\bibitem [{\citenamefont {Yang}\ \emph {et~al.}(2013)\citenamefont {Yang},
  \citenamefont {Cui},\ and\ \citenamefont {DiSalvo}}]{Yang2013}%
  \BibitemOpen
  \bibfield  {author} {\bibinfo {author} {\bibfnamefont {M.}~\bibnamefont
  {Yang}}, \bibinfo {author} {\bibfnamefont {Z.}~\bibnamefont {Cui}}, \ and\
  \bibinfo {author} {\bibfnamefont {F.~J.}\ \bibnamefont {DiSalvo}},\ }\href
  {\doibase 10.1039/c3cp51109j} {\bibfield  {journal} {\bibinfo  {journal}
  {Phys. Chem. Chem. Phys.}\ }\textbf {\bibinfo {volume} {15}},\ \bibinfo
  {pages} {7041} (\bibinfo {year} {2013})}\BibitemShut {NoStop}%
\bibitem [{\citenamefont {Gall}\ \emph {et~al.}(2002)\citenamefont {Gall},
  \citenamefont {Shin}, \citenamefont {Haasch}, \citenamefont {Petrov},\ and\
  \citenamefont {Greene}}]{Gall2002}%
  \BibitemOpen
  \bibfield  {author} {\bibinfo {author} {\bibfnamefont {D.}~\bibnamefont
  {Gall}}, \bibinfo {author} {\bibfnamefont {C.-S.}\ \bibnamefont {Shin}},
  \bibinfo {author} {\bibfnamefont {R.~T.}\ \bibnamefont {Haasch}}, \bibinfo
  {author} {\bibfnamefont {I.}~\bibnamefont {Petrov}}, \ and\ \bibinfo {author}
  {\bibfnamefont {J.~E.}\ \bibnamefont {Greene}},\ }\href {\doibase
  10.1063/1.1466528} {\bibfield  {journal} {\bibinfo  {journal} {J. Appl.
  Phys.}\ }\textbf {\bibinfo {volume} {91}},\ \bibinfo {pages} {5882} (\bibinfo
  {year} {2002})}\BibitemShut {NoStop}%
\bibitem [{\citenamefont {Mientus}\ and\ \citenamefont
  {Ellmer}(1999)}]{Mientus1999}%
  \BibitemOpen
  \bibfield  {author} {\bibinfo {author} {\bibfnamefont {R.}~\bibnamefont
  {Mientus}}\ and\ \bibinfo {author} {\bibfnamefont {K.}~\bibnamefont
  {Ellmer}},\ }\href {\doibase 10.1016/S0257-8972(99)00124-3} {\bibfield
  {journal} {\bibinfo  {journal} {Surf. Coat. Tech.}\ }\textbf {\bibinfo
  {volume} {116-119}},\ \bibinfo {pages} {1093} (\bibinfo {year}
  {1999})}\BibitemShut {NoStop}%
\bibitem [{\citenamefont {Shih}(1986)}]{Shih1986}%
  \BibitemOpen
  \bibfield  {author} {\bibinfo {author} {\bibfnamefont {K.~K.}\ \bibnamefont
  {Shih}},\ }\href {\doibase 10.1116/1.573887} {\bibfield  {journal} {\bibinfo
  {journal} {J. Vac. Sci. Tech. A}\ }\textbf {\bibinfo {volume} {4}},\ \bibinfo
  {pages} {564} (\bibinfo {year} {1986})}\BibitemShut {NoStop}%
\bibitem [{\citenamefont {Constantin}\ \emph {et~al.}(2004)\citenamefont
  {Constantin}, \citenamefont {Haider}, \citenamefont {Ingram},\ and\
  \citenamefont {Smith}}]{Constantin2004}%
  \BibitemOpen
  \bibfield  {author} {\bibinfo {author} {\bibfnamefont {C.}~\bibnamefont
  {Constantin}}, \bibinfo {author} {\bibfnamefont {M.~B.}\ \bibnamefont
  {Haider}}, \bibinfo {author} {\bibfnamefont {D.}~\bibnamefont {Ingram}}, \
  and\ \bibinfo {author} {\bibfnamefont {A.~R.}\ \bibnamefont {Smith}},\ }\href
  {\doibase 10.1063/1.1836878} {\bibfield  {journal} {\bibinfo  {journal}
  {Appl. Phys. Lett.}\ }\textbf {\bibinfo {volume} {85}},\ \bibinfo {pages}
  {6371} (\bibinfo {year} {2004})}\BibitemShut {NoStop}%
\bibitem [{\citenamefont {Inumaru}\ \emph {et~al.}(2007)\citenamefont
  {Inumaru}, \citenamefont {Koyama}, \citenamefont {Imo-oka},\ and\
  \citenamefont {Yamanaka}}]{Inumaru2007}%
  \BibitemOpen
  \bibfield  {author} {\bibinfo {author} {\bibfnamefont {K.}~\bibnamefont
  {Inumaru}}, \bibinfo {author} {\bibfnamefont {K.}~\bibnamefont {Koyama}},
  \bibinfo {author} {\bibfnamefont {N.}~\bibnamefont {Imo-oka}}, \ and\
  \bibinfo {author} {\bibfnamefont {S.}~\bibnamefont {Yamanaka}},\ }\href
  {\doibase 10.1103/PhysRevB.75.054416} {\bibfield  {journal} {\bibinfo
  {journal} {Phys. Rev. B}\ }\textbf {\bibinfo {volume} {75}},\ \bibinfo
  {pages} {054416} (\bibinfo {year} {2007})}\BibitemShut {NoStop}%
\bibitem [{\citenamefont {Browne}\ \emph {et~al.}(1970)\citenamefont {Browne},
  \citenamefont {Liddell}, \citenamefont {Street},\ and\ \citenamefont
  {Mills}}]{Browne1970}%
  \BibitemOpen
  \bibfield  {author} {\bibinfo {author} {\bibfnamefont {J.~D.}\ \bibnamefont
  {Browne}}, \bibinfo {author} {\bibfnamefont {P.~R.}\ \bibnamefont {Liddell}},
  \bibinfo {author} {\bibfnamefont {R.}~\bibnamefont {Street}}, \ and\ \bibinfo
  {author} {\bibfnamefont {T.}~\bibnamefont {Mills}},\ }\href {\doibase
  10.1002/pssa.19700010411} {\bibfield  {journal} {\bibinfo  {journal} {Phys.
  Status Solidi A}\ }\textbf {\bibinfo {volume} {1}},\ \bibinfo {pages} {715}
  (\bibinfo {year} {1970})}\BibitemShut {NoStop}%
\bibitem [{\citenamefont {Herle}\ \emph {et~al.}(1997)\citenamefont {Herle},
  \citenamefont {Hegde}, \citenamefont {Vasathacharya}, \citenamefont {Philip},
  \citenamefont {Rama~Rao},\ and\ \citenamefont
  {Sripathi}}]{SubramanyaHerle1997}%
  \BibitemOpen
  \bibfield  {author} {\bibinfo {author} {\bibfnamefont {P.~S.}\ \bibnamefont
  {Herle}}, \bibinfo {author} {\bibfnamefont {M.}~\bibnamefont {Hegde}},
  \bibinfo {author} {\bibfnamefont {N.}~\bibnamefont {Vasathacharya}}, \bibinfo
  {author} {\bibfnamefont {S.}~\bibnamefont {Philip}}, \bibinfo {author}
  {\bibfnamefont {M.}~\bibnamefont {Rama~Rao}}, \ and\ \bibinfo {author}
  {\bibfnamefont {T.}~\bibnamefont {Sripathi}},\ }\href {\doibase
  10.1006/jssc.1997.7554} {\bibfield  {journal} {\bibinfo  {journal} {J. Solid
  State Chem.}\ }\textbf {\bibinfo {volume} {134}},\ \bibinfo {pages} {120}
  (\bibinfo {year} {1997})}\BibitemShut {NoStop}%
\bibitem [{\citenamefont {Quintela}\ \emph {et~al.}(2009)\citenamefont
  {Quintela}, \citenamefont {Rivadulla},\ and\ \citenamefont
  {Rivas}}]{Quintela2009ThermoelectricCrN}%
  \BibitemOpen
  \bibfield  {author} {\bibinfo {author} {\bibfnamefont {C.~X.}\ \bibnamefont
  {Quintela}}, \bibinfo {author} {\bibfnamefont {F.}~\bibnamefont {Rivadulla}},
  \ and\ \bibinfo {author} {\bibfnamefont {J.}~\bibnamefont {Rivas}},\ }\href
  {\doibase 10.1063/1.3120280} {\bibfield  {journal} {\bibinfo  {journal}
  {Appl. Phys. Lett.}\ }\textbf {\bibinfo {volume} {94}},\ \bibinfo {pages}
  {152103} (\bibinfo {year} {2009})}\BibitemShut {NoStop}%
\bibitem [{\citenamefont {Alam}\ \emph {et~al.}(2017)\citenamefont {Alam},
  \citenamefont {Disseler}, \citenamefont {Ratcliff}, \citenamefont {Borchers},
  \citenamefont {Ponce-Perez}, \citenamefont {Cocoletzi}, \citenamefont
  {Takeuchi}, \citenamefont {Foley}, \citenamefont {Richard}, \citenamefont
  {Ingram},\ and\ \citenamefont {Smith}}]{Alam2017}%
  \BibitemOpen
  \bibfield  {author} {\bibinfo {author} {\bibfnamefont {K.}~\bibnamefont
  {Alam}}, \bibinfo {author} {\bibfnamefont {S.~M.}\ \bibnamefont {Disseler}},
  \bibinfo {author} {\bibfnamefont {W.~D.}\ \bibnamefont {Ratcliff}}, \bibinfo
  {author} {\bibfnamefont {J.~A.}\ \bibnamefont {Borchers}}, \bibinfo {author}
  {\bibfnamefont {R.}~\bibnamefont {Ponce-Perez}}, \bibinfo {author}
  {\bibfnamefont {G.~H.}\ \bibnamefont {Cocoletzi}}, \bibinfo {author}
  {\bibfnamefont {N.}~\bibnamefont {Takeuchi}}, \bibinfo {author}
  {\bibfnamefont {A.}~\bibnamefont {Foley}}, \bibinfo {author} {\bibfnamefont
  {A.}~\bibnamefont {Richard}}, \bibinfo {author} {\bibfnamefont {D.~C.}\
  \bibnamefont {Ingram}}, \ and\ \bibinfo {author} {\bibfnamefont {A.~R.}\
  \bibnamefont {Smith}},\ }\href@noop {} {} (\bibinfo {year} {2017}),\ \Eprint
  {http://arxiv.org/abs/arXiv:1703.03829} {arXiv:1703.03829} \BibitemShut
  {NoStop}%
\bibitem [{\citenamefont {Filippetti}\ \emph {et~al.}(1999)\citenamefont
  {Filippetti}, \citenamefont {Pickett},\ and\ \citenamefont
  {Klein}}]{Filippetti1999}%
  \BibitemOpen
  \bibfield  {author} {\bibinfo {author} {\bibfnamefont {A.}~\bibnamefont
  {Filippetti}}, \bibinfo {author} {\bibfnamefont {W.~E.}\ \bibnamefont
  {Pickett}}, \ and\ \bibinfo {author} {\bibfnamefont {B.~M.}\ \bibnamefont
  {Klein}},\ }\href {\doibase 10.1103/PhysRevB.59.7043} {\bibfield  {journal}
  {\bibinfo  {journal} {Phys. Rev. B}\ }\textbf {\bibinfo {volume} {59}},\
  \bibinfo {pages} {7043} (\bibinfo {year} {1999})}\BibitemShut {NoStop}%
\bibitem [{\citenamefont {Filippetti}\ and\ \citenamefont
  {Hill}(2000)}]{Filippetti2000}%
  \BibitemOpen
  \bibfield  {author} {\bibinfo {author} {\bibfnamefont {A.}~\bibnamefont
  {Filippetti}}\ and\ \bibinfo {author} {\bibfnamefont {N.}~\bibnamefont
  {Hill}},\ }\href {\doibase 10.1103/PhysRevLett.85.5166} {\bibfield  {journal}
  {\bibinfo  {journal} {Phys. Rev. Lett.}\ }\textbf {\bibinfo {volume} {85}},\
  \bibinfo {pages} {5166} (\bibinfo {year} {2000})}\BibitemShut {NoStop}%
\bibitem [{\citenamefont {Botana}\ \emph {et~al.}(2012)\citenamefont {Botana},
  \citenamefont {Tran}, \citenamefont {Pardo}, \citenamefont {Baldomir},\ and\
  \citenamefont {Blaha}}]{Botana2012}%
  \BibitemOpen
  \bibfield  {author} {\bibinfo {author} {\bibfnamefont {A.~S.}\ \bibnamefont
  {Botana}}, \bibinfo {author} {\bibfnamefont {F.}~\bibnamefont {Tran}},
  \bibinfo {author} {\bibfnamefont {V.}~\bibnamefont {Pardo}}, \bibinfo
  {author} {\bibfnamefont {D.}~\bibnamefont {Baldomir}}, \ and\ \bibinfo
  {author} {\bibfnamefont {P.}~\bibnamefont {Blaha}},\ }\href {\doibase
  10.1103/PhysRevB.85.235118} {\bibfield  {journal} {\bibinfo  {journal} {Phys.
  Rev. B}\ }\textbf {\bibinfo {volume} {85}},\ \bibinfo {pages} {235118}
  (\bibinfo {year} {2012})}\BibitemShut {NoStop}%
\bibitem [{\citenamefont {Zhou}\ \emph {et~al.}(2014)\citenamefont {Zhou},
  \citenamefont {K{\"{o}}rmann}, \citenamefont {Holec}, \citenamefont
  {Bartosik}, \citenamefont {Grabowski}, \citenamefont {Neugebauer},\ and\
  \citenamefont {Mayrhofer}}]{Zhou2014}%
  \BibitemOpen
  \bibfield  {author} {\bibinfo {author} {\bibfnamefont {L.}~\bibnamefont
  {Zhou}}, \bibinfo {author} {\bibfnamefont {F.}~\bibnamefont {K{\"{o}}rmann}},
  \bibinfo {author} {\bibfnamefont {D.}~\bibnamefont {Holec}}, \bibinfo
  {author} {\bibfnamefont {M.}~\bibnamefont {Bartosik}}, \bibinfo {author}
  {\bibfnamefont {B.}~\bibnamefont {Grabowski}}, \bibinfo {author}
  {\bibfnamefont {J.}~\bibnamefont {Neugebauer}}, \ and\ \bibinfo {author}
  {\bibfnamefont {P.~H.}\ \bibnamefont {Mayrhofer}},\ }\href {\doibase
  10.1103/PhysRevB.90.184102} {\bibfield  {journal} {\bibinfo  {journal} {Phys.
  Rev. B}\ }\textbf {\bibinfo {volume} {90}},\ \bibinfo {pages} {184102}
  (\bibinfo {year} {2014})}\BibitemShut {NoStop}%
\bibitem [{\citenamefont {Ikeyama}\ \emph {et~al.}(2016)\citenamefont
  {Ikeyama}, \citenamefont {Suzuki}, \citenamefont {Suzuki}, \citenamefont
  {Nakayama}, \citenamefont {Suematsu},\ and\ \citenamefont
  {Niihara}}]{Ikeyama2016a}%
  \BibitemOpen
  \bibfield  {author} {\bibinfo {author} {\bibfnamefont {S.}~\bibnamefont
  {Ikeyama}}, \bibinfo {author} {\bibfnamefont {K.}~\bibnamefont {Suzuki}},
  \bibinfo {author} {\bibfnamefont {T.}~\bibnamefont {Suzuki}}, \bibinfo
  {author} {\bibfnamefont {T.}~\bibnamefont {Nakayama}}, \bibinfo {author}
  {\bibfnamefont {H.}~\bibnamefont {Suematsu}}, \ and\ \bibinfo {author}
  {\bibfnamefont {K.}~\bibnamefont {Niihara}},\ }\href {\doibase
  10.7567/JJAP.55.02BC02} {\bibfield  {journal} {\bibinfo  {journal} {Jap. J.
  Appl. Phys.}\ }\textbf {\bibinfo {volume} {55}},\ \bibinfo {pages} {02BC02}
  (\bibinfo {year} {2016})}\BibitemShut {NoStop}%
\bibitem [{\citenamefont {Zhang}\ \emph {et~al.}(2014)\citenamefont {Zhang},
  \citenamefont {Li}, \citenamefont {Zhao},\ and\ \citenamefont
  {Wang}}]{Zhang2014a}%
  \BibitemOpen
  \bibfield  {author} {\bibinfo {author} {\bibfnamefont {S.}~\bibnamefont
  {Zhang}}, \bibinfo {author} {\bibfnamefont {Y.}~\bibnamefont {Li}}, \bibinfo
  {author} {\bibfnamefont {T.}~\bibnamefont {Zhao}}, \ and\ \bibinfo {author}
  {\bibfnamefont {Q.}~\bibnamefont {Wang}},\ }\href {\doibase
  10.1038/srep05241} {\bibfield  {journal} {\bibinfo  {journal} {Sci. Rep.}\
  }\textbf {\bibinfo {volume} {4}},\ \bibinfo {pages} {5241} (\bibinfo {year}
  {2014})}\BibitemShut {NoStop}%
\bibitem [{\citenamefont {Conley}\ \emph {et~al.}(2013)\citenamefont {Conley},
  \citenamefont {Wang}, \citenamefont {Ziegler}, \citenamefont {Haglund},
  \citenamefont {Pantelides},\ and\ \citenamefont
  {Bolotin}}]{BolotinStrain2013}%
  \BibitemOpen
  \bibfield  {author} {\bibinfo {author} {\bibfnamefont {H.~J.}\ \bibnamefont
  {Conley}}, \bibinfo {author} {\bibfnamefont {B.}~\bibnamefont {Wang}},
  \bibinfo {author} {\bibfnamefont {J.~I.}\ \bibnamefont {Ziegler}}, \bibinfo
  {author} {\bibfnamefont {R.~F.}\ \bibnamefont {Haglund}}, \bibinfo {author}
  {\bibfnamefont {S.~T.}\ \bibnamefont {Pantelides}}, \ and\ \bibinfo {author}
  {\bibfnamefont {K.~I.}\ \bibnamefont {Bolotin}},\ }\href {\doibase
  10.1021/nl4014748} {\bibfield  {journal} {\bibinfo  {journal} {Nano Lett.}\
  }\textbf {\bibinfo {volume} {13}},\ \bibinfo {pages} {3626} (\bibinfo {year}
  {2013})}\BibitemShut {NoStop}%
\bibitem [{\citenamefont {Giannozzi}\ \emph {et~al.}(2009)\citenamefont
  {Giannozzi}, \citenamefont {Baroni}, \citenamefont {Bonini}, \citenamefont
  {Calandra}, \citenamefont {Car}, \citenamefont {Cavazzoni}, \citenamefont
  {Ceresoli}, \citenamefont {Chiarotti}, \citenamefont {Cococcioni},
  \citenamefont {Dabo}, \citenamefont {Dal~Corso}, \citenamefont
  {de~Gironcoli}, \citenamefont {Fabris}, \citenamefont {Fratesi},
  \citenamefont {Gebauer}, \citenamefont {Gerstmann}, \citenamefont
  {Gougoussis}, \citenamefont {Kokalj}, \citenamefont {Lazzeri}, \citenamefont
  {Martin-Samos}, \citenamefont {Marzari}, \citenamefont {Mauri}, \citenamefont
  {Mazzarello}, \citenamefont {Paolini}, \citenamefont {Pasquarello},
  \citenamefont {Paulatto}, \citenamefont {Sbraccia}, \citenamefont {Scandolo},
  \citenamefont {Sclauzero}, \citenamefont {Seitsonen}, \citenamefont
  {Smogunov}, \citenamefont {Umari},\ and\ \citenamefont
  {Wentzcovitch}}]{Giannozzi2009}%
  \BibitemOpen
  \bibfield  {author} {\bibinfo {author} {\bibfnamefont {P.}~\bibnamefont
  {Giannozzi}}, \bibinfo {author} {\bibfnamefont {S.}~\bibnamefont {Baroni}},
  \bibinfo {author} {\bibfnamefont {N.}~\bibnamefont {Bonini}}, \bibinfo
  {author} {\bibfnamefont {M.}~\bibnamefont {Calandra}}, \bibinfo {author}
  {\bibfnamefont {R.}~\bibnamefont {Car}}, \bibinfo {author} {\bibfnamefont
  {C.}~\bibnamefont {Cavazzoni}}, \bibinfo {author} {\bibfnamefont
  {D.}~\bibnamefont {Ceresoli}}, \bibinfo {author} {\bibfnamefont {G.~L.}\
  \bibnamefont {Chiarotti}}, \bibinfo {author} {\bibfnamefont {M.}~\bibnamefont
  {Cococcioni}}, \bibinfo {author} {\bibfnamefont {I.}~\bibnamefont {Dabo}},
  \bibinfo {author} {\bibfnamefont {A.}~\bibnamefont {Dal~Corso}}, \bibinfo
  {author} {\bibfnamefont {S.}~\bibnamefont {de~Gironcoli}}, \bibinfo {author}
  {\bibfnamefont {S.}~\bibnamefont {Fabris}}, \bibinfo {author} {\bibfnamefont
  {G.}~\bibnamefont {Fratesi}}, \bibinfo {author} {\bibfnamefont
  {R.}~\bibnamefont {Gebauer}}, \bibinfo {author} {\bibfnamefont
  {U.}~\bibnamefont {Gerstmann}}, \bibinfo {author} {\bibfnamefont
  {C.}~\bibnamefont {Gougoussis}}, \bibinfo {author} {\bibfnamefont
  {A.}~\bibnamefont {Kokalj}}, \bibinfo {author} {\bibfnamefont
  {M.}~\bibnamefont {Lazzeri}}, \bibinfo {author} {\bibfnamefont
  {L.}~\bibnamefont {Martin-Samos}}, \bibinfo {author} {\bibfnamefont
  {N.}~\bibnamefont {Marzari}}, \bibinfo {author} {\bibfnamefont
  {F.}~\bibnamefont {Mauri}}, \bibinfo {author} {\bibfnamefont
  {R.}~\bibnamefont {Mazzarello}}, \bibinfo {author} {\bibfnamefont
  {S.}~\bibnamefont {Paolini}}, \bibinfo {author} {\bibfnamefont
  {A.}~\bibnamefont {Pasquarello}}, \bibinfo {author} {\bibfnamefont
  {L.}~\bibnamefont {Paulatto}}, \bibinfo {author} {\bibfnamefont
  {C.}~\bibnamefont {Sbraccia}}, \bibinfo {author} {\bibfnamefont
  {S.}~\bibnamefont {Scandolo}}, \bibinfo {author} {\bibfnamefont
  {G.}~\bibnamefont {Sclauzero}}, \bibinfo {author} {\bibfnamefont {A.~P.}\
  \bibnamefont {Seitsonen}}, \bibinfo {author} {\bibfnamefont {A.}~\bibnamefont
  {Smogunov}}, \bibinfo {author} {\bibfnamefont {P.}~\bibnamefont {Umari}}, \
  and\ \bibinfo {author} {\bibfnamefont {R.~M.}\ \bibnamefont {Wentzcovitch}},\
  }\href {\doibase 10.1088/0953-8984/21/39/395502} {\bibfield  {journal}
  {\bibinfo  {journal} {J. Phys.: Cond. Matt.}\ }\textbf {\bibinfo {volume}
  {21}},\ \bibinfo {pages} {395502} (\bibinfo {year} {2009})}\BibitemShut
  {NoStop}%
\bibitem [{\citenamefont {Anisimov}\ \emph {et~al.}(1997)\citenamefont
  {Anisimov}, \citenamefont {Aryasetiawan},\ and\ \citenamefont
  {Lichtenstein}}]{Anisimov1997}%
  \BibitemOpen
  \bibfield  {author} {\bibinfo {author} {\bibfnamefont {V.~I.}\ \bibnamefont
  {Anisimov}}, \bibinfo {author} {\bibfnamefont {F.}~\bibnamefont
  {Aryasetiawan}}, \ and\ \bibinfo {author} {\bibfnamefont {A.~I.}\
  \bibnamefont {Lichtenstein}},\ }\href {\doibase 10.1088/0953-8984/9/4/002}
  {\bibfield  {journal} {\bibinfo  {journal} {J. Phys.: Cond. Matt.}\ }\textbf
  {\bibinfo {volume} {9}},\ \bibinfo {pages} {767} (\bibinfo {year}
  {1997})}\BibitemShut {NoStop}%
\bibitem [{\citenamefont {Liechtenstein}\ \emph {et~al.}(1995)\citenamefont
  {Liechtenstein}, \citenamefont {Anisimov},\ and\ \citenamefont
  {Zaanen}}]{Liechtenstein1995}%
  \BibitemOpen
  \bibfield  {author} {\bibinfo {author} {\bibfnamefont {A.~I.}\ \bibnamefont
  {Liechtenstein}}, \bibinfo {author} {\bibfnamefont {V.~I.}\ \bibnamefont
  {Anisimov}}, \ and\ \bibinfo {author} {\bibfnamefont {J.}~\bibnamefont
  {Zaanen}},\ }\href {\doibase 10.1103/PhysRevB.52.R5467} {\bibfield  {journal}
  {\bibinfo  {journal} {Phys. Rev. B}\ }\textbf {\bibinfo {volume} {52}},\
  \bibinfo {pages} {R5467} (\bibinfo {year} {1995})}\BibitemShut {NoStop}%
\bibitem [{\citenamefont {Herwadkar}\ and\ \citenamefont
  {Lambrecht}(2009)}]{Herwadkar2009}%
  \BibitemOpen
  \bibfield  {author} {\bibinfo {author} {\bibfnamefont {A.}~\bibnamefont
  {Herwadkar}}\ and\ \bibinfo {author} {\bibfnamefont {W.}~\bibnamefont
  {Lambrecht}},\ }\href {\doibase 10.1103/PhysRevB.79.035125} {\bibfield
  {journal} {\bibinfo  {journal} {Phys. Rev. B}\ }\textbf {\bibinfo {volume}
  {79}},\ \bibinfo {pages} {035125} (\bibinfo {year} {2009})}\BibitemShut
  {NoStop}%
\bibitem [{\citenamefont {Cococcioni}\ and\ \citenamefont
  {de~Gironcoli}(2005)}]{Cococcioni2005}%
  \BibitemOpen
  \bibfield  {author} {\bibinfo {author} {\bibfnamefont {M.}~\bibnamefont
  {Cococcioni}}\ and\ \bibinfo {author} {\bibfnamefont {S.}~\bibnamefont
  {de~Gironcoli}},\ }\href {\doibase 10.1103/PhysRevB.71.035105} {\bibfield
  {journal} {\bibinfo  {journal} {Phys. Rev. B}\ }\textbf {\bibinfo {volume}
  {71}},\ \bibinfo {pages} {035105} (\bibinfo {year} {2005})}\BibitemShut
  {NoStop}%
\bibitem [{\citenamefont {Wright}(1997)}]{Wright1997}%
  \BibitemOpen
  \bibfield  {author} {\bibinfo {author} {\bibfnamefont {A.~F.}\ \bibnamefont
  {Wright}},\ }\href {\doibase 10.1063/1.366114} {\bibfield  {journal}
  {\bibinfo  {journal} {J. Appl. Phys.}\ }\textbf {\bibinfo {volume} {82}},\
  \bibinfo {pages} {2833} (\bibinfo {year} {1997})}\BibitemShut {NoStop}%
\bibitem [{\citenamefont {Murnaghan}(1944)}]{Murnaghan1944}%
  \BibitemOpen
  \bibfield  {author} {\bibinfo {author} {\bibfnamefont {F.~D.}\ \bibnamefont
  {Murnaghan}},\ }\href {\doibase 10.1073/pnas.30.9.244} {\bibfield  {journal}
  {\bibinfo  {journal} {PNAS}\ }\textbf {\bibinfo {volume} {30}},\ \bibinfo
  {pages} {244} (\bibinfo {year} {1944})}\BibitemShut {NoStop}%
\bibitem [{\citenamefont {Antonov}\ and\ \citenamefont
  {Iordanova}(2010)}]{Antonov2010}%
  \BibitemOpen
  \bibfield  {author} {\bibinfo {author} {\bibfnamefont {V.}~\bibnamefont
  {Antonov}}\ and\ \bibinfo {author} {\bibfnamefont {I.}~\bibnamefont
  {Iordanova}},\ }in\ \href {\doibase 10.1063/1.3322328} {\emph {\bibinfo
  {booktitle} {AIP Conf. Proc.}}},\ Vol.\ \bibinfo {volume} {1203}\ (\bibinfo
  {year} {2010})\ pp.\ \bibinfo {pages} {1149--1154}\BibitemShut {NoStop}%
\bibitem [{\citenamefont {Liang}\ \emph {et~al.}(2010)\citenamefont {Liang},
  \citenamefont {Yuan},\ and\ \citenamefont {Zhang}}]{Liang2010}%
  \BibitemOpen
  \bibfield  {author} {\bibinfo {author} {\bibfnamefont {Y.}~\bibnamefont
  {Liang}}, \bibinfo {author} {\bibfnamefont {X.}~\bibnamefont {Yuan}}, \ and\
  \bibinfo {author} {\bibfnamefont {W.}~\bibnamefont {Zhang}},\ }\href
  {\doibase 10.1016/j.ssc.2010.08.005} {\bibfield  {journal} {\bibinfo
  {journal} {Solid State Commun.}\ }\textbf {\bibinfo {volume} {150}},\
  \bibinfo {pages} {2045} (\bibinfo {year} {2010})}\BibitemShut {NoStop}%
\bibitem [{\citenamefont {Jiao}\ \emph {et~al.}(2013)\citenamefont {Jiao},
  \citenamefont {Niu}, \citenamefont {Ma},\ and\ \citenamefont
  {Huang}}]{JIAO2013}%
  \BibitemOpen
  \bibfield  {author} {\bibinfo {author} {\bibfnamefont {Z.-Y.}\ \bibnamefont
  {Jiao}}, \bibinfo {author} {\bibfnamefont {Y.-J.}\ \bibnamefont {Niu}},
  \bibinfo {author} {\bibfnamefont {S.-H.}\ \bibnamefont {Ma}}, \ and\ \bibinfo
  {author} {\bibfnamefont {X.-F.}\ \bibnamefont {Huang}},\ }\href {\doibase
  10.1142/S0217984913501583} {\bibfield  {journal} {\bibinfo  {journal} {Mod.
  Phys. Lett. B}\ }\textbf {\bibinfo {volume} {27}},\ \bibinfo {pages}
  {1350158} (\bibinfo {year} {2013})}\BibitemShut {NoStop}%
\bibitem [{\citenamefont {Almer}\ \emph {et~al.}(2003)\citenamefont {Almer},
  \citenamefont {Lienert}, \citenamefont {Peng}, \citenamefont {Schlauer},\
  and\ \citenamefont {Od{\'{e}}n}}]{Almer2003}%
  \BibitemOpen
  \bibfield  {author} {\bibinfo {author} {\bibfnamefont {J.}~\bibnamefont
  {Almer}}, \bibinfo {author} {\bibfnamefont {U.}~\bibnamefont {Lienert}},
  \bibinfo {author} {\bibfnamefont {R.~L.}\ \bibnamefont {Peng}}, \bibinfo
  {author} {\bibfnamefont {C.}~\bibnamefont {Schlauer}}, \ and\ \bibinfo
  {author} {\bibfnamefont {M.}~\bibnamefont {Od{\'{e}}n}},\ }\href {\doibase
  10.1063/1.1582351} {\bibfield  {journal} {\bibinfo  {journal} {J. Appl.
  Phys.}\ }\textbf {\bibinfo {volume} {94}},\ \bibinfo {pages} {697} (\bibinfo
  {year} {2003})}\BibitemShut {NoStop}%
\bibitem [{\citenamefont {Martinschitz}\ \emph {et~al.}(2009)\citenamefont
  {Martinschitz}, \citenamefont {Daniel}, \citenamefont {Mitterer},\ and\
  \citenamefont {Keckes}}]{Martinschitz2009}%
  \BibitemOpen
  \bibfield  {author} {\bibinfo {author} {\bibfnamefont {K.~J.}\ \bibnamefont
  {Martinschitz}}, \bibinfo {author} {\bibfnamefont {R.}~\bibnamefont
  {Daniel}}, \bibinfo {author} {\bibfnamefont {C.}~\bibnamefont {Mitterer}}, \
  and\ \bibinfo {author} {\bibfnamefont {J.}~\bibnamefont {Keckes}},\ }\href
  {\doibase 10.1107/S0021889809011807} {\bibfield  {journal} {\bibinfo
  {journal} {J. Appl. Crystallogr.}\ }\textbf {\bibinfo {volume} {42}},\
  \bibinfo {pages} {416} (\bibinfo {year} {2009})}\BibitemShut {NoStop}%
\bibitem [{\citenamefont {Muir}\ \emph {et~al.}(1987)\citenamefont {Muir},
  \citenamefont {Perz},\ and\ \citenamefont {Fawcett}}]{Muir1987}%
  \BibitemOpen
  \bibfield  {author} {\bibinfo {author} {\bibfnamefont {W.~C.}\ \bibnamefont
  {Muir}}, \bibinfo {author} {\bibfnamefont {J.~M.}\ \bibnamefont {Perz}}, \
  and\ \bibinfo {author} {\bibfnamefont {E.}~\bibnamefont {Fawcett}},\ }\href
  {\doibase 10.1088/0305-4608/17/12/017} {\bibfield  {journal} {\bibinfo
  {journal} {J. Phys. F: Met. Phys.}\ }\textbf {\bibinfo {volume} {17}},\
  \bibinfo {pages} {2431} (\bibinfo {year} {1987})}\BibitemShut {NoStop}%
\bibitem [{\citenamefont {Qteish}\ \emph {et~al.}(2006)\citenamefont {Qteish},
  \citenamefont {Rinke}, \citenamefont {Scheffler},\ and\ \citenamefont
  {Neugebauer}}]{Qteish2006}%
  \BibitemOpen
  \bibfield  {author} {\bibinfo {author} {\bibfnamefont {A.}~\bibnamefont
  {Qteish}}, \bibinfo {author} {\bibfnamefont {P.}~\bibnamefont {Rinke}},
  \bibinfo {author} {\bibfnamefont {M.}~\bibnamefont {Scheffler}}, \ and\
  \bibinfo {author} {\bibfnamefont {J.}~\bibnamefont {Neugebauer}},\ }\href
  {\doibase 10.1103/PhysRevB.74.245208} {\bibfield  {journal} {\bibinfo
  {journal} {Phys. Rev. B}\ }\textbf {\bibinfo {volume} {74}},\ \bibinfo
  {pages} {1} (\bibinfo {year} {2006})}\BibitemShut {NoStop}%
\bibitem [{\citenamefont {Dhar}\ \emph {et~al.}(2007)\citenamefont {Dhar},
  \citenamefont {Ungersb{\"{o}}ck}, \citenamefont {Kosina}, \citenamefont
  {Grasser},\ and\ \citenamefont {Selberherr}}]{Dhar2007}%
  \BibitemOpen
  \bibfield  {author} {\bibinfo {author} {\bibfnamefont {S.}~\bibnamefont
  {Dhar}}, \bibinfo {author} {\bibfnamefont {E.}~\bibnamefont
  {Ungersb{\"{o}}ck}}, \bibinfo {author} {\bibfnamefont {H.}~\bibnamefont
  {Kosina}}, \bibinfo {author} {\bibfnamefont {T.}~\bibnamefont {Grasser}}, \
  and\ \bibinfo {author} {\bibfnamefont {S.}~\bibnamefont {Selberherr}},\
  }\href {\doibase 10.1109/TNANO.2006.888533} {\bibfield  {journal} {\bibinfo
  {journal} {IEEE Trans. Nanotech.}\ }\textbf {\bibinfo {volume} {6}},\
  \bibinfo {pages} {97} (\bibinfo {year} {2007})}\BibitemShut {NoStop}%
\bibitem [{\citenamefont {Kim}\ \emph {et~al.}(1997)\citenamefont {Kim},
  \citenamefont {Lambrecht},\ and\ \citenamefont {Segall}}]{Kim1997}%
  \BibitemOpen
  \bibfield  {author} {\bibinfo {author} {\bibfnamefont {K.}~\bibnamefont
  {Kim}}, \bibinfo {author} {\bibfnamefont {W.}~\bibnamefont {Lambrecht}}, \
  and\ \bibinfo {author} {\bibfnamefont {B.}~\bibnamefont {Segall}},\ }\href
  {\doibase 10.1103/PhysRevB.56.7018.2} {\bibfield  {journal} {\bibinfo
  {journal} {Phys. Rev. B}\ }\textbf {\bibinfo {volume} {56}},\ \bibinfo
  {pages} {7018} (\bibinfo {year} {1997})}\BibitemShut {NoStop}%
\bibitem [{\citenamefont {Qin}\ \emph {et~al.}(2013)\citenamefont {Qin},
  \citenamefont {Duan}, \citenamefont {Shi}, \citenamefont {Shi},\ and\
  \citenamefont {Tang}}]{QinStrains}%
  \BibitemOpen
  \bibfield  {author} {\bibinfo {author} {\bibfnamefont {L.}~\bibnamefont
  {Qin}}, \bibinfo {author} {\bibfnamefont {Y.}~\bibnamefont {Duan}}, \bibinfo
  {author} {\bibfnamefont {H.}~\bibnamefont {Shi}}, \bibinfo {author}
  {\bibfnamefont {L.}~\bibnamefont {Shi}}, \ and\ \bibinfo {author}
  {\bibfnamefont {G.}~\bibnamefont {Tang}},\ }\href
  {http://stacks.iop.org/0953-8984/25/i=4/a=045801} {\bibfield  {journal}
  {\bibinfo  {journal} {J. Phys.: Cond. Matt.}\ }\textbf {\bibinfo {volume}
  {25}},\ \bibinfo {pages} {045801} (\bibinfo {year} {2013})}\BibitemShut
  {NoStop}%
\bibitem [{\citenamefont {Bhobe}\ \emph {et~al.}(2010)\citenamefont {Bhobe},
  \citenamefont {Chainani}, \citenamefont {Taguchi}, \citenamefont {Takeuchi},
  \citenamefont {Eguchi}, \citenamefont {Matsunami}, \citenamefont {Ishizaka},
  \citenamefont {Takata}, \citenamefont {Oura}, \citenamefont {Senba},
  \citenamefont {Ohashi}, \citenamefont {Nishino}, \citenamefont {Yabashi},
  \citenamefont {Tamasaku}, \citenamefont {Ishikawa}, \citenamefont {Takenaka},
  \citenamefont {Takagi},\ and\ \citenamefont {Shin}}]{Bhobe2010}%
  \BibitemOpen
  \bibfield  {author} {\bibinfo {author} {\bibfnamefont {P.~A.}\ \bibnamefont
  {Bhobe}}, \bibinfo {author} {\bibfnamefont {A.}~\bibnamefont {Chainani}},
  \bibinfo {author} {\bibfnamefont {M.}~\bibnamefont {Taguchi}}, \bibinfo
  {author} {\bibfnamefont {T.}~\bibnamefont {Takeuchi}}, \bibinfo {author}
  {\bibfnamefont {R.}~\bibnamefont {Eguchi}}, \bibinfo {author} {\bibfnamefont
  {M.}~\bibnamefont {Matsunami}}, \bibinfo {author} {\bibfnamefont
  {K.}~\bibnamefont {Ishizaka}}, \bibinfo {author} {\bibfnamefont
  {Y.}~\bibnamefont {Takata}}, \bibinfo {author} {\bibfnamefont
  {M.}~\bibnamefont {Oura}}, \bibinfo {author} {\bibfnamefont {Y.}~\bibnamefont
  {Senba}}, \bibinfo {author} {\bibfnamefont {H.}~\bibnamefont {Ohashi}},
  \bibinfo {author} {\bibfnamefont {Y.}~\bibnamefont {Nishino}}, \bibinfo
  {author} {\bibfnamefont {M.}~\bibnamefont {Yabashi}}, \bibinfo {author}
  {\bibfnamefont {K.}~\bibnamefont {Tamasaku}}, \bibinfo {author}
  {\bibfnamefont {T.}~\bibnamefont {Ishikawa}}, \bibinfo {author}
  {\bibfnamefont {K.}~\bibnamefont {Takenaka}}, \bibinfo {author}
  {\bibfnamefont {H.}~\bibnamefont {Takagi}}, \ and\ \bibinfo {author}
  {\bibfnamefont {S.}~\bibnamefont {Shin}},\ }\href {\doibase
  10.1103/PhysRevLett.104.236404} {\bibfield  {journal} {\bibinfo  {journal}
  {Phys. Rev. Lett.}\ }\textbf {\bibinfo {volume} {104}},\ \bibinfo {pages}
  {236404} (\bibinfo {year} {2010})}\BibitemShut {NoStop}%
\end{thebibliography}%
\end{document}